\documentclass[aps,prb,twocolumn,floats,showpacs,superscriptaddress]{revtex4-1}
\usepackage{graphicx,epsfig}
\usepackage{times,bbm}
\usepackage{graphics,dcolumn,bm,float}
\usepackage{amssymb,amsmath,rotate,color}
\usepackage[title,titletoc,toc]{appendix}
\usepackage[pagebackref=false,colorlinks,linkcolor=blue,citecolor=blue,urlcolor=blue]{hyperref}

\begin{document}
    \unitlength 1 cm
    \newcommand{\etal}{{\em et. al.}}
    \newcommand{\be}{\begin{equation}}
    \newcommand{\ee}{\end{equation}}
    \newcommand{\bearr}{\begin{eqnarray}}
    \newcommand{\eearr}{\end{eqnarray}}
    \newcommand{\nn}{\nonumber}
    \newcommand{\la}{\langle}
    \newcommand{\ra}{\rangle}
    \newcommand{\cd}{c^\dagger}
    \newcommand{\vd}{v^\dagger}
    \newcommand{\ad}{a^\dagger}
    \newcommand{\bd}{b^\dagger}
    \newcommand{\tk}{{\tilde{k}}}
    \newcommand{\tp}{{\tilde{p}}}
    \newcommand{\tq}{{\tilde{q}}}
    \newcommand{\eps}{\varepsilon}
    \newcommand{\vk}{{\vec k}}
    \newcommand{\vp}{{\vec p}}
    \newcommand{\vq}{{\vec q}}
    \newcommand{\vkp}{\vec {k'}}
    \newcommand{\vpp}{\vec {p'}}
    \newcommand{\vqp}{\vec {q'}}
    \newcommand{\bk}{{\bf k}}
    \newcommand{\bp}{{\bf p}}
    \newcommand{\bq}{{\bf q}}
    \newcommand{\br}{{\bf r}}
    \newcommand{\bR}{{\bf R}}
    \newcommand{\up}{\uparrow}
    \newcommand{\down}{\downarrow}
    \newcommand{\fns}{\footnotesize}
    \newcommand{\ns}{\normalsize}
    \newcommand{\cdag}{c^{\dagger}}
    \newcommand{\lc}{\langle\!\langle}
    \newcommand{\rc}{\rangle\!\rangle}

    \title{Resilience of Majorana Fermions in the face of Disorder}
    \author{Alireza Habibi} 
    \affiliation{Department of Physics, Sharif University of Technology, Tehran 11155-9161, Iran}
    \author{S. A. Jafari}
    \email{akbar.jafari@gmail.com}
    \affiliation{Department of Physics, Sharif University of Technology, Tehran 11155-9161, Iran}
    \author{S. Rouhani}
    \affiliation{Department of Physics, Sharif University of Technology, Tehran 11155-9161, Iran}
    \affiliation{Institute for Research in Fundamental Sciences (IPM), Tehran 19395-5531, Iran}
    \date{\today}
    
    \begin{abstract} 
We elucidate the reduction of the winding number (WN) caused by the onsite disorder in a higher WN next nearest neighbor XY model. 
When disorder becomes strong enough, Majorana edge modes become critically extended, beyond which they collapse into Anderson localized (AL) 
states in the bulk, resulting in a topological Anderson insulating state (TAI). We identify a resilience threshold $W_t$ for every pair of 
Majorana fermions (MFs). In response to increasing disorder every pair of MFs collapse into AL bulk at their resilience threshold. 
For very strong disorder, all Majorana fermions collapse and a topologically trivial state is obtained. 
We show that the threshold values are deeply related to the localization length of Majorana fermions, which can be efficiently calculated by an appropriate modification of the transfer matrix method. At the topological transition point, 
localization length of the zero modes diverges and the system becomes scale invariant. 
The number of peaks in the localization length as the function of disorder strength determines the number of zero modes in the clean state before disorder is introduced. This finding elevates the transfer matrix method to the level of a tool
for determination of the topological index of both clean and disordered systems. 
    \end{abstract} 
    \maketitle 
    
\section{Introduction}
Generically topology protects the system against weak disorder. However, when disorder can close the gap, it can change topological properties~\cite{Altland2015}. Therefore when disorder becomes strong, it can affect the topology and can lead to a quantum phase transition to the so-called topological Anderson insulating (TAI) state~\cite{Jian2009}. The interplay between disorder and topology has been studied in the past~\cite{Jian2009}. The change of Hall conductance as a function of disorder in two-dimensional square lattice was numerically investigated in Ref.~\onlinecite{Hatsugai1999}. In the presence of a disorder that breaks the symmetries but preserves them in ensemble averages, the system is still strictly characterized by topological numbers~\cite{Prodan2015}, although they may be different numbers compared to clean system. In the vicinity of topological Anderson transition, the Chern number in finite-size two-dimensional systems smoothly changes across the two topological phases. For an infinite system, the change becomes a sharp step function~\cite{Resta2011,Zhang2013}. 
Kitaev chain with a quasi-periodic potential with Fibonacci sequence shows an interplay of fractal and topology for Fibonacci potentials in the topological phase diagram\cite{ghadimi2009}. In a two-dimensional bipartite lattice, topological number survives up to a much stronger disorder (one order of magnitude higher) if only one of the sub-lattices is disordered and the final state in the strongly disordered system is in the metallic phase~\cite{Vozmediano2015}.
But what is the mechanism by which the topology changes by adding disorder to the system?

Quadratic Hamiltonians have been classified by Altland and Zirnbauer into ten classes~\cite{Altland1997} based on symmetry classes of random matrices~\cite{SchnyderRMP2016}. In every space dimension, five of the ten symmetry classes of  Altland-Zirnbauer classification have non-trivial topology~\cite{SchnyderRMP2016}. In 1D systems, these five topologically non-trivial classes are D,  DIII which are classified with a $Z_2$ index and three chiral classes AIII,  BDI, and CII which have the $Z$ classification. 
Recently a generalization of XY spin chain~\cite{Martin2014} has been proposed which after fermionization corresponds to BDI class of topological superconductors~\cite{Jafari2016}. This model can be classified with an integer WN. This extension called nXY in Ref.~\onlinecite{Jafari2016} allows for engineering a Hamiltonian with any desired WN. Therefore it provides a playground to study systems with larger WNs and hence larger number of Majorana fermions.  
A similar extension of Ising in transverse field Hamiltonian exists~\cite{Sen}. 
     In this work, we study the effect of disorder on the topological properties of the second neighbor generalization of the XY model (2XY model)~\cite{Jafari2016}
\bearr
   H_{\rm 2XY}=&\sum_{j}(J_1+\lambda_1)\sigma^x_j\sigma^x_{j+1}+(J_1-\lambda_1)\sigma^y_j\sigma^y_{j+1}\\
    &+(J_2+\lambda_2)\sigma^x_j\sigma^z_{j+1}\sigma^x_{j+2}+(J_2-\lambda_2)\sigma^y_j\sigma^z_{j+1}\sigma^y_{j+2},\nn
    \label{2xy.eqn}
\eearr
     which after Jordan-Wigner (JW) fermionization becomes,
    \bearr
    H_{\rm 2XY}^{\rm JW}&=2\sum_{s=1,2}\sum_{j} {J_s c^\dagger_j c_{j+s}+\lambda_s c_j^\dagger c_{j+s}^\dagger}
    + {\rm h.c.}
    \label{2xyJW.eqn}
    \eearr
This Hamiltonian allows for WNs up to two. 
In this class, any short-range correlated disorder is enough to cause Anderson localization of all quasi-particles~\cite{Vishveshwara2001}. For open boundary conditions, there can exist some symmetry protected (Majorana) zero modes which are localized on the boundary~\cite{Kitaev-topo}.  Strong disorder at transition points will change localized topological states to Anderson localized states. This transition was dubbed topological Anderson transition~\cite{Jian2009}.

\section{Model Hamiltonian} 
\label{model}

The Hamiltonian we study is given by 
\be
   H=H_{\rm 2XY}+H_{\rm dis}
\ee
where the translation-invariant part $H_{\rm 2XY}$ of the model is given by Eq.~(\ref{2xy.eqn}).
The random-field term $H_{\rm dis}$ is of the following form,
    \begin{align}
    H_{\rm dis}&=\sum_j{\left(\varepsilon_j +\mu\right)\sigma_j^z}.
    \end{align}
where $\varepsilon_j$ is uniformly distributed in the interval $\left[ -W/2, W/2\right]$ where $W$ is onsite disorder strength.

The clean system can be solved with JW transformation~\cite{Jafari2016}:
    \begin{align}
    \sigma_j^z=2c_j^\dagger c_j-1,
    \sigma_j^x=e^{i \phi_j} (c_j^\dagger+ c_j),
    \sigma_j^x=i e^{i \phi_j} (c_j^\dagger- c_j),
    \end{align}
    where $ \phi_j=\pi \sum_{l<j}{c_l^\dagger c_l} $ is phase string. With this transformation
    the entire Hamiltonian becomes,
    \begin{multline}
    H=2\sum_{s=1,r}\sum_{j} {J_s c^\dagger_j c_{j+s}+\lambda_s c_j^\dagger c_{j+s}^\dagger}+ {\rm h.c.}\\
    +2\sum_{j} {\left(\varepsilon_j
        +\mu\right) \left(c^\dagger_j c_{j}-\frac{1}{2}\right)}
    \end{multline}
    This makes it clear that $\mu$ is chemical potential of the JW fermions. 
    This model is an extension of Kitaev chain model of a p-wave superconductor where 
    next nearest neighbors hoppings and pairings along with onsite disorder have been added.

    In terms of Nambu spionors $\psi^\dagger=\begin{pmatrix} c^\dagger & c\end{pmatrix}$ we have:
    \begin{align}
    H= \psi^\dagger \begin{pmatrix} \tilde H_0 && \tilde\Delta \\ \tilde \Delta ^\dagger  && -\tilde H_0 \end{pmatrix}\psi
    \equiv \psi^\dagger \tilde H \psi,
    \end{align}
or equivalently introducing Pauli matrices $\vec\tau$ for the Nambu space the 
matrix representation of the Hamiltonian becomes,
    \begin{align}
    \tilde H=\tilde H_0\otimes \tau_z +i \tilde\Delta \otimes \tau_y 
    \end{align}
    where $\tilde H_0$ and $\tilde \Delta$ are the following matrices,
    \begin{align}
    \tilde H_0&=\sum_{s=1,r}\sum_{j} {
        (J_s |j\rangle \langle {j+s}| + h.c)}+\sum_{j} {\left(\varepsilon_j
        +\mu\right)( |j\rangle \langle {j}|)},\\
    \tilde \Delta&=\sum_{s=1,r}\sum_{j}{
        (\lambda_s |j\rangle_e \langle{j+s}|_h -\lambda_s|j+s\rangle_e \langle j|_h)}.
    \end{align}
    Note that since the matrix $\tilde\Delta$ is off-diagonal in the Nambu space,
    the projection operators used in definition of $\tilde\Delta$ are indeed of the $|\rangle_e \langle |_h$ form.
    Apparently we have $\Delta^\dagger=-\Delta$. The $\tilde H_0$ part is diagonal in the Nambu space
    and hence $e$ or $h$ subscript is not necessary. The operator for the particle-hole
    symmetry can be defined as $S=\mathbbm{1}\otimes\tau_x$ where $\mathbbm{1}$ acts on the space of site 
    indices $j=1,\ldots,N$ and $\tau_x$ acts on the Nambu space which simply replaces $c$ and $c^\dagger$ and can be seen to affect the
    Hamiltonian as,
    \begin{align}
    S^{-1}H S=-H.
    \end{align}
This model belongs to BDI class of topological superconductors~\cite{Jafari2016}. This is exactly solvable and
in section~\ref{exactwn.sec}, we obtain a closed form formula for the winding number of the clean system which varies between $-2$ and $+2$.

\section{Exact WN for clean 2XY model} 
\label{exactwn.sec}
Let us add an arbitrary chemical potential to Eq.~\eqref{2xyJW.eqn} which gives 
an extension of the Kitaev model of topological superconductor~\cite{Kitaev-topo} as follows,
    \begin{multline}
    H=2\sum_{j} \sum_{s=1,2}{J_s c^\dagger_j c_{j+s}+\lambda_s c_j^\dagger c_{j+s}^\dagger}  + {\rm h.c.}\\
    +2\sum_{j}{\mu (c^\dagger_j c_{j}-\frac{1}{2})}
    \end{multline}
    This Hamiltonian can be rewritten in terms of Majorana fermions $ a_{i}=c_i^\dagger+c_i$ and
$b_{i}=i(c_i^\dagger-c_i)$ which obey commutation rules:
    \begin{align}
    \{a_{i},a_{j}\}=\{b_{i},b_{j}\}=2\delta_{i,j},~~~~~~\{a_{i},b_{j}\}=0 
    \end{align}
    and furthermore are self-adjoint, $ a_{i}^\dagger =a_{i}$, $ b_{i}^\dagger =b_{i}$. 
    The Hamiltonian in terms of Majorana fermions becomes,
    \begin{multline}
    H=i\sum_{s=1,2}\sum_{j} \left( J_s-\lambda_s\right) a_{j}b_{j+s}
    +\left( -J_s-\lambda_s\right) b_{j}a_{j+s} \\
    +\frac{\mu}{2} \left(a_{j}b_{j}- b_{j}a_{j}\right).
    \end{multline}
\begin{figure}[b]
    \centering
    \includegraphics[width=\linewidth]{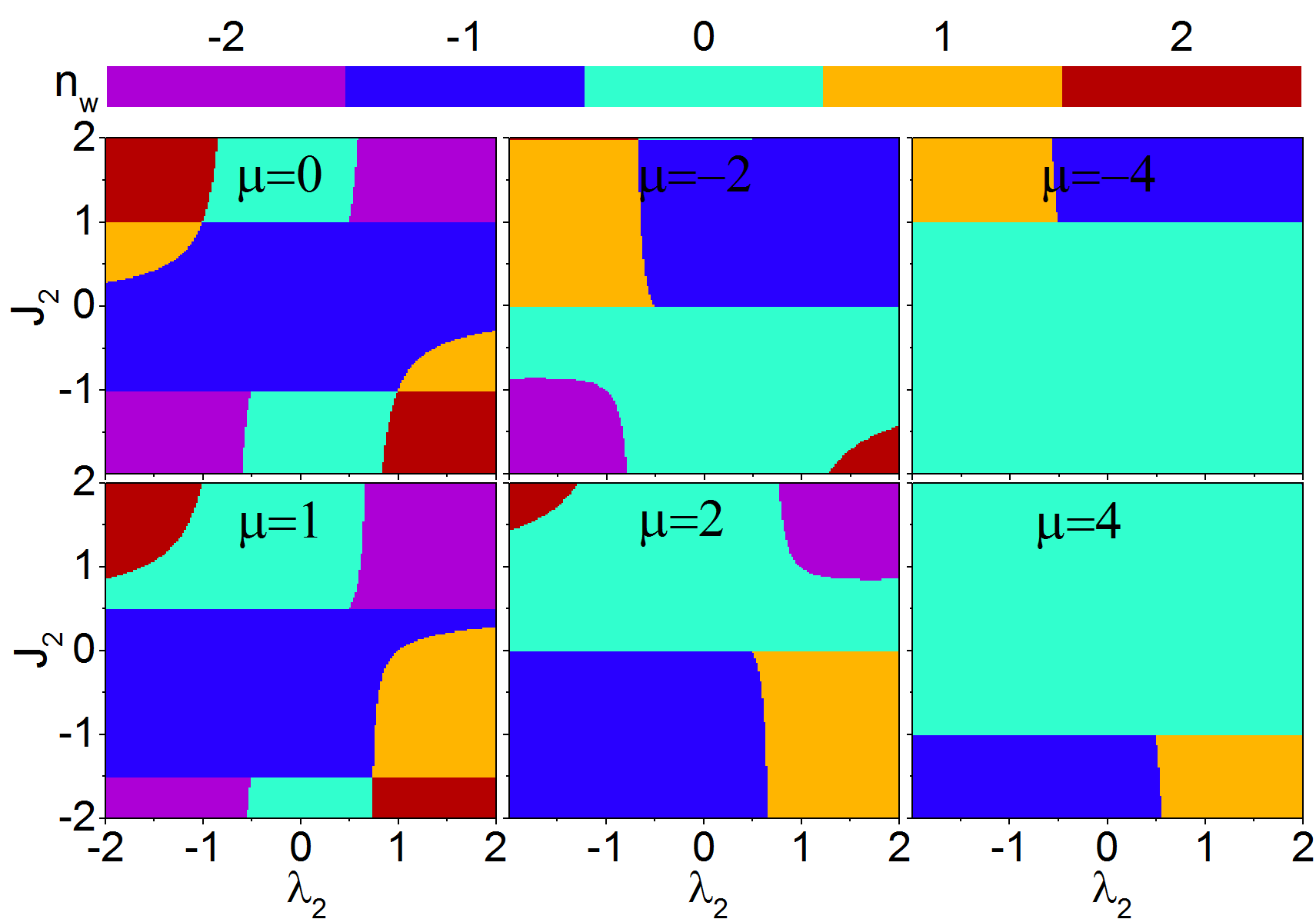}
    \caption{Map of the WN for the clean 2XY model for various values of chemical potential}
    \label{fig:mu0}
\end{figure}
    It can be represented in the $k$-space as, 
    \begin{align}
    H=\sum_{k\in BZ} 
    \begin{pmatrix}
    a_{k} & b_{k} 
    \end{pmatrix}
    \begin{pmatrix} 
    0 &h(k)  \\
    h(k)^*   & 0 
    \end{pmatrix}
    \begin{pmatrix}
    a_{-k} \\
    b_{-k}
    \end{pmatrix},
    \end{align}
    where,
    \begin{align}
    \frac{h(k)}{2}=
    i\left( J_1 \cos k+ J_2 \cos2k+\dfrac{\mu}{2}\right) 
    &+\left( \lambda_1 \sin k+ \lambda_2 \sin2k \right).
    \end{align}
    This Hamiltonian is  invariant under time reversal symmetry and particle-hole symmetry and belongs to class BDI~\cite{Jafari2016}. 
    This canonical form allows us to calculate the WN as,~\cite{Prodan-Non-commutative}
    \begin{align}
    n_w=\frac{1}{2\pi}\Im{\int_{-\pi}^\pi{dk h(k)^{-1}\partial_k h(k)}} .
    \end{align}
    For $ h(k) =r_k e^{i \theta_k}$ one gets, $ n_w=\frac{1}{2\pi}\int_{-\pi}^\pi{dk \partial_k \theta_k}=n$ where $ n \in  \mathbb{Z}$ and 
    \begin{align}
    \theta_k=\mbox{arctan} \frac{J_1 \cos k+J_2 \cos 2k+\dfrac{\mu}{2}}{\lambda_1\sin k+\lambda_2 \sin 2k}.
    \label{atn.eqn}
    \end{align}
    The $\theta_k$ is {\em not} continues for  $k$ in range $ \left[-\pi , \pi \right]$.
    When the argument of arctan in Eq.~\eqref{atn.eqn} diverges, $\theta$ becomes discontinuous. 
    To correctly count the total phase winding, one has to exclude such singular points.  
    Therefore we break the integration to,
    \begin{align}
    n_w=\sum_{P_i}\frac{1}{2\pi}\int_{P_i^+}^{P_{i+1}^-}{dk \partial_k \theta_k}
    =\frac{1}{2\pi}\sum_{ i=0,N-1}\theta_k \Big|_{P_i^+}^{P_{i+1}^-},
    \end{align}
    where $ P_0=-\pi $ and $ P_N=\pi $ and other $ P_i, i=1,...,N-1 $ are singular points of Eq.~\eqref{atn.eqn}.  
    In the present model, we have three singular points $\theta_k$ at 
    $P_1= \arccos \frac{-\lambda_1}{2\lambda_2}$ , $P_2= \pi-\arccos \frac{-\lambda_1}{2\lambda_2}$ and $ P_3=0 $.
    Obviously the points $P_1$ and $P_2$ are present as long as,
    $|\lambda_1| \leqslant |2\lambda_2|$. 
    With these provisions, straightforward algebra gives,  
    \begin{widetext}
    \begin{multline}
    n_w=-\frac{1}{2}
    \text{sgn}\left[\left(-J_1+J_2+\dfrac{\mu}{2}\right) \left(-\lambda _1+2 \lambda _2\right)\right] 
    -\frac{1}{2}\text{sgn}\left[\left(+J_1+J_2+\dfrac{\mu}{2}\right) \left(+\lambda _1+2 \lambda _2\right)\right]\\
    - \Theta \left(1-\left| \frac{\lambda _1}{2 \lambda _2}\right| \right) 
    \times \text{sgn}\left[-\frac{\lambda _1^2}{2 \lambda _2}+2 \lambda _2 \left(\frac{\lambda _1^2}{2 \lambda _2^2}-1\right)\right] 
    \times  \text{sgn}\left[-\frac{J_1 \lambda _1}{2 \lambda _2}+J_2 \left(\frac{\lambda _1^2}{2 \lambda _2^2}-1\right)+\dfrac{\mu}{2} \right].
    \end{multline}
    \end{widetext}
    Here $\Theta$ is the Heaviside function. The allowed WNs in this model are  $ n_w=-2,-1,0,1,2 $
    which is plotted in Fig.~\ref{fig:mu0} for various values of the chemical potential $\mu$ as indicated in the legend. 
\begin{figure}[b]
    \centering
    \includegraphics[width=\linewidth]{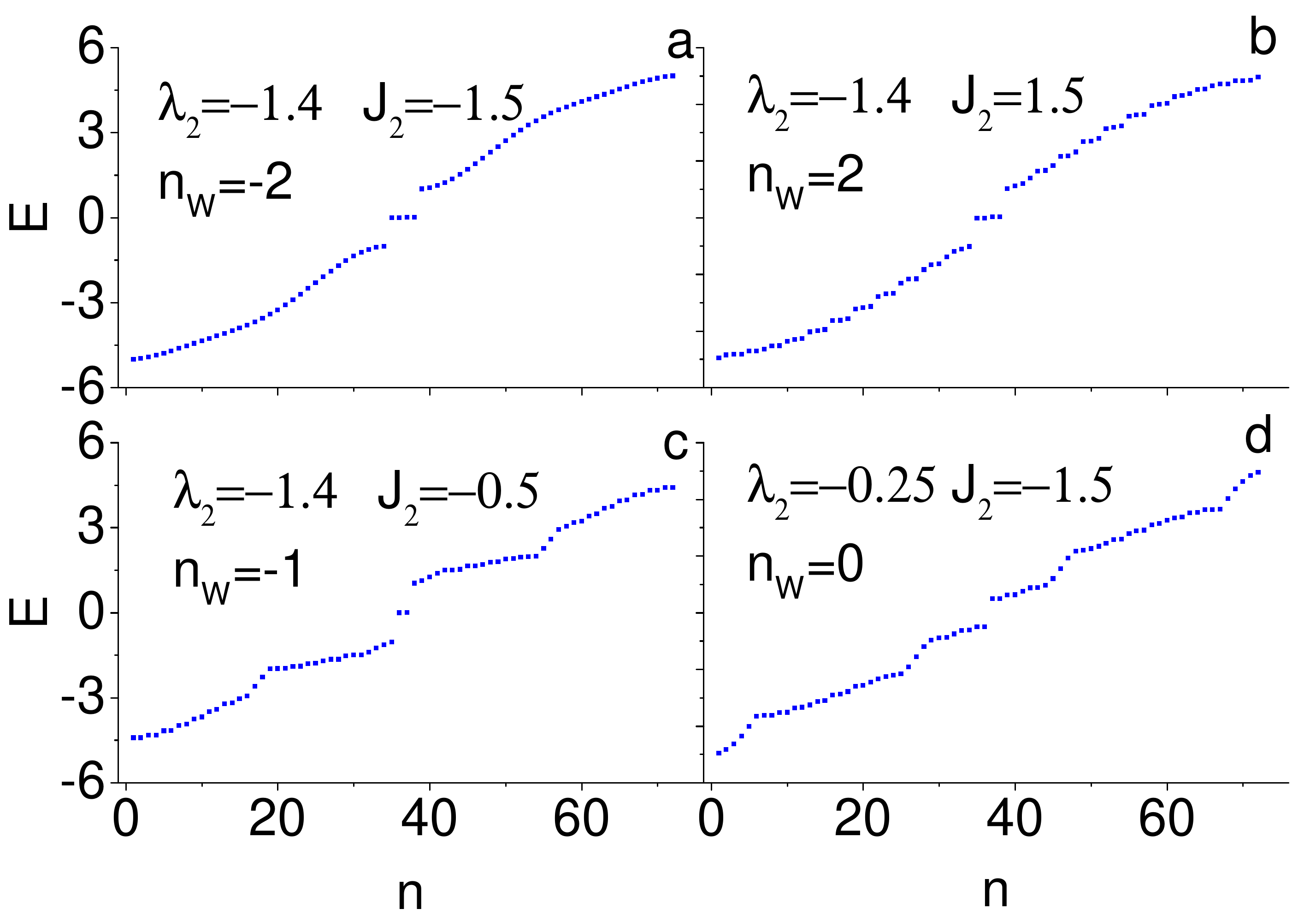}
    \caption{Energy eigenvalues for different values of $ \lambda_2 $ and $  J_2 $. For open boundary condition 
    we have $4$, $2$ and $0$ zero energy Majorana modes for different values of $ \lambda_2 $ and $  J_2 $. }
    \label{fig:eig}
\end{figure}
For zero chemical potential (corresponding to particle-hole symmetry)
the phase diagram is given by the top left
panel of Fig.~\ref{fig:mu0}. This is in agreement with Ref.~\onlinecite{Jafari2016}. 
When adding the disorder term $H_{\rm dis}$ we will be interested in $\mu=0$ case where the system enjoys a particle-hole symmetry 
Note that we will assume $J_1=1 and \lambda_2=1$.  
At $\mu=0$, one can also do exact diagonalization on a small $L=40$ system
to see the correspondence between the winding number and the number of MF pairs. In Fig.~\ref{fig:eig}
we show the number of zero energy states and read the WN read from the top left panel of Fig.~\ref{fig:mu0}.
As can be seen, the sign of winding number does not matter and the number of Majorana zero modes
are given by $2|n_w|$, meaning that every WN corresponds to one {\em pair} of MFs. The $a$-type MF is localized in one end, while the
partner MF of $b$-type is localized in the other end. Sign reversal of WN simply swaps the $a$ and $b$
partners across the chain~\cite{Jafari2016}. 
\begin{figure}[b]
\centering
\includegraphics[width=0.99\linewidth]{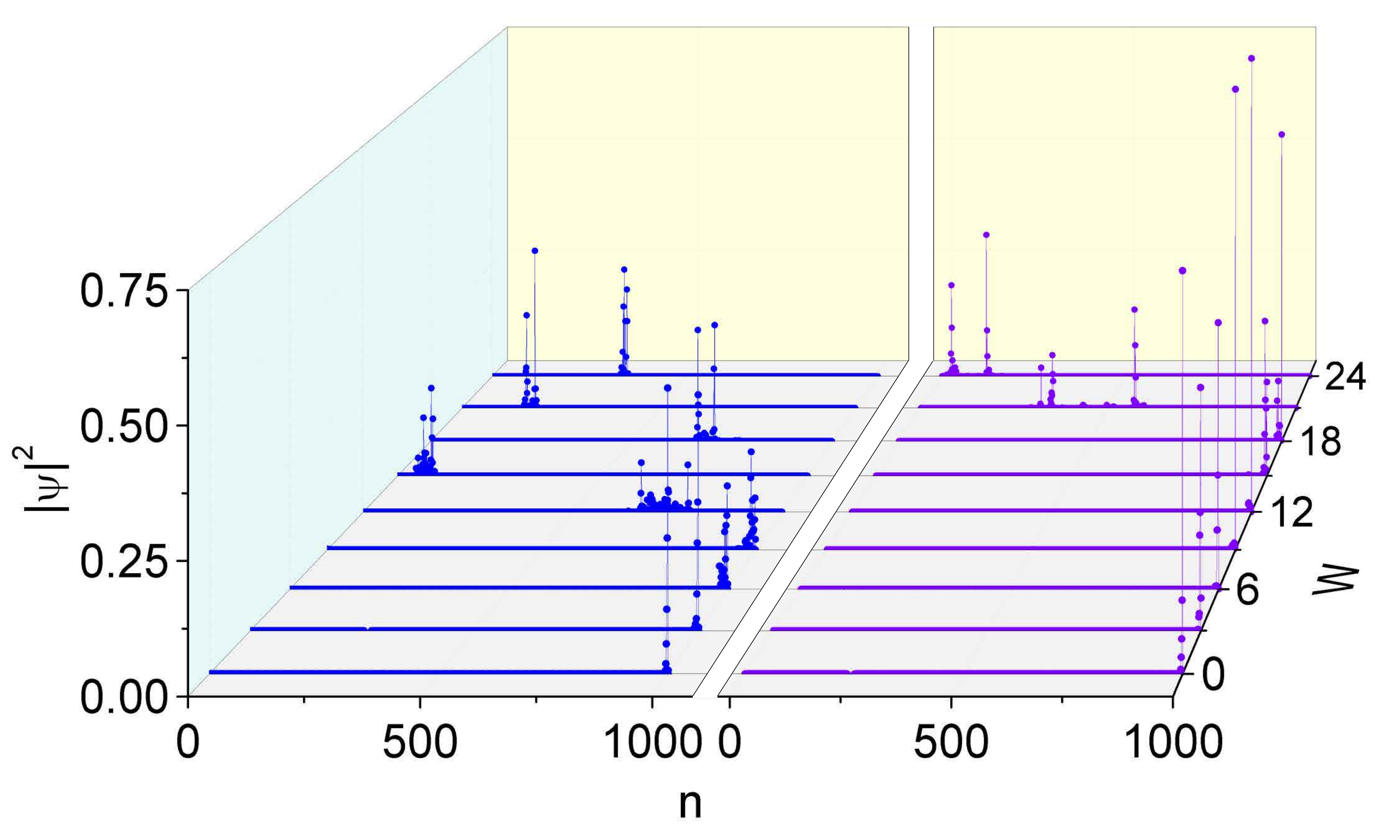} 
\caption{Zero energy wave functions for two type of zero modes for $L=10^3$ sites.  
The first pair of MFs become Anderson localized (AL) upon crossing $W_{t1}$ (left) and the second pair are AL after crossing $W_{t2}$.
}
\label{psi_mf}
\end{figure}

\section {Indications of resilience}
As shown in the Fig.~\ref{fig:eig}, each WN (irrespective of its sign) corresponds a pair of zero modes localized at two ends of the system. 
The sign of WN can be changed by simply renaming $a$ and $b$ MFs that compose a fermion. 
Now let us add the on-site Anderson disorder $\sum_i\varepsilon_i c^\dagger_i c_i$ to clean 2XY
Hamiltonian~\eqref{2xyJW.eqn}, where the onsite energies are uniformly distributed in a range of
width $W$. Unless otherwise specified, we will do most of the analysis for a representative 
parameter values $J_2=-1.5,\lambda_2=-1.4$, corresponding to $n_w=-2$, where there are two pairs of MFs.

\subsection{Spectral manifestations}
Let us start by examining the spectrum. 
In Fig.~\ref{psi_mf} we present the square of zero-energy eigenfunctions 
for $L=10^3$ sites. We focus on those MFs that for small values of disorder are localized in the right edge of the system. By increasing disorder, at a first
threshold value, $W_{t1}\approx 10.5$ the first pair of MFs are depinned from the edge,
while the second mode still persists and remains edge localized. Eventually, beyond
a larger threshold, $W_{t2}$ the second pair of MFs are also Anderson localized into the bulk. 

\begin{figure}[t]
\centering
\includegraphics[width=0.495\linewidth]{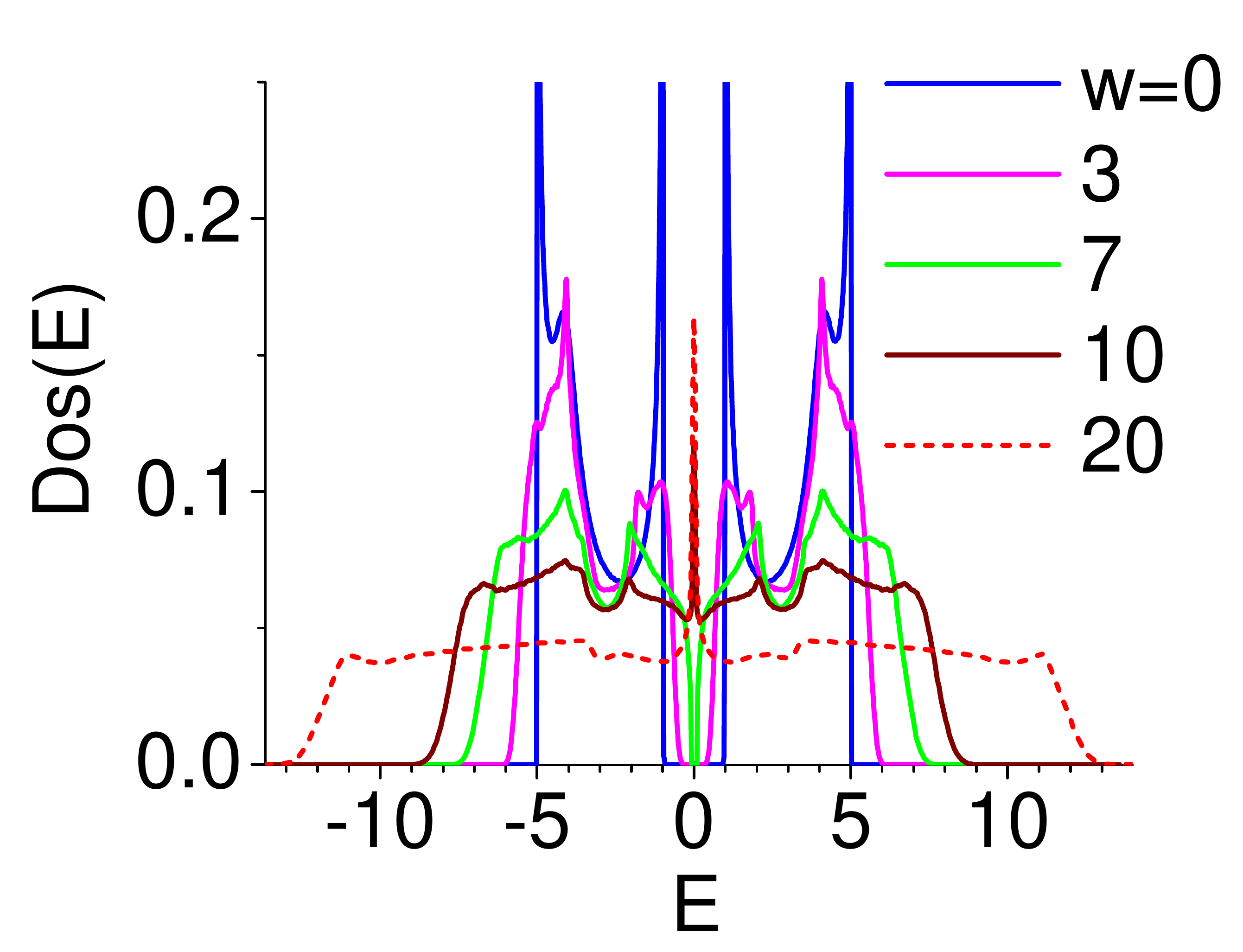}
\includegraphics[width=0.495\linewidth]{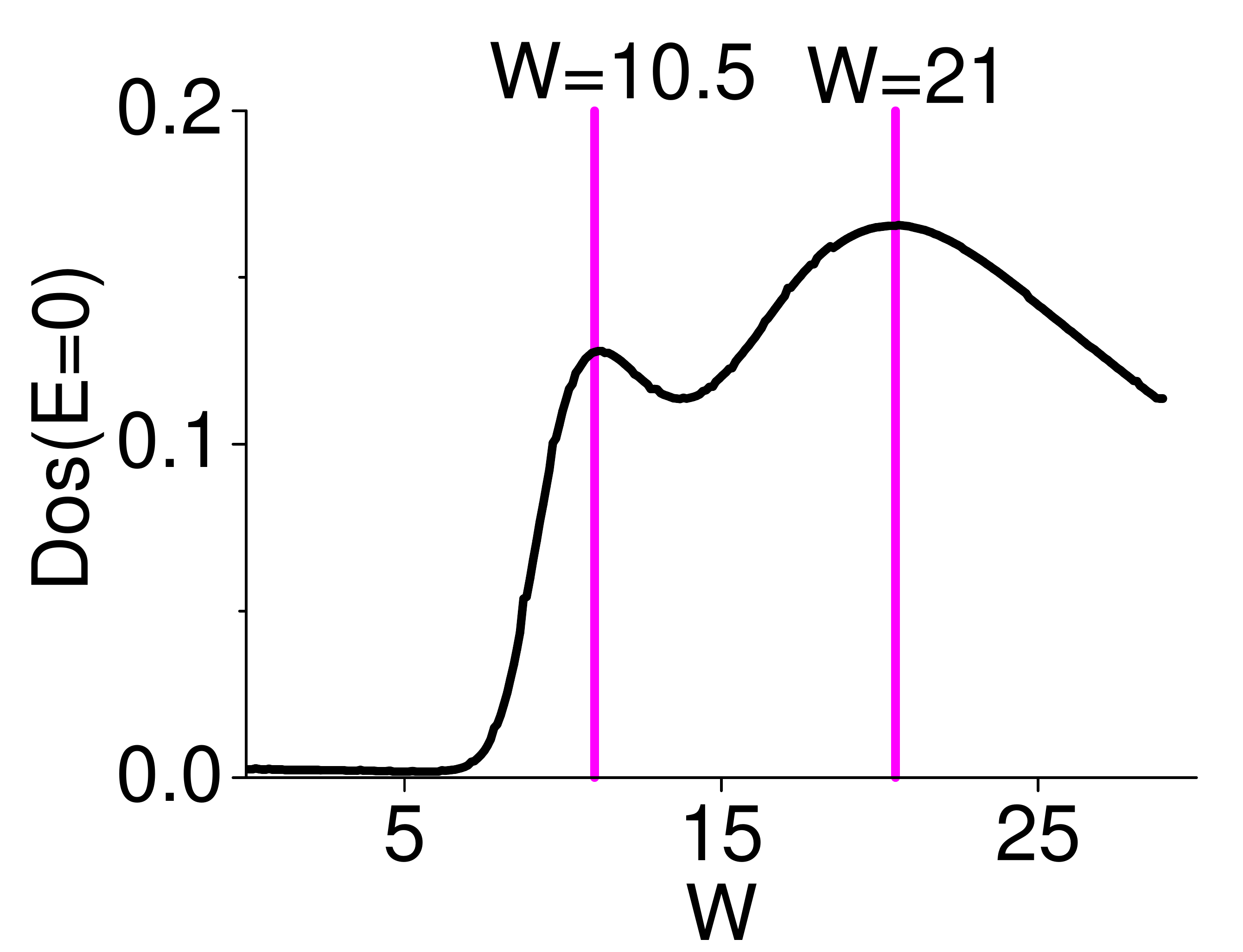}
\caption{(left) The DOS for disordered system obtained with kernel polynomial method (appendix~\ref{kpm.sec})
Due to level repulsion, when disorder increases, the gap will be filled. 
(right) The DOS $\rho_0$ at the middle of the spectrum, $E=0$ as a function of disorder strength, $W$.
The resilience threshold shows up as a maximum in $\rho_0$ versus $W$ curve. 
}
\label{fig:E-Dos}
\end{figure}

To see what is happening in the Hilbert space, in Fig.~\ref{fig:E-Dos} we have plotted the density of states (DOS)
for various values of the disorder. The left panel indicates that by increasing disorder, the clean superconducting gap of the spectrum is gradually filled by the level repulsion mechanism. In the right panel which is obtained by kernel
polynomial method (KPM)~\cite{Lee1,Amini-Jafari-2009,Habibi,Fehske} we plot the DOS $\rho_0$ at $E=0$ as a function of disorder strength, $W$.
In this method (see appendix~\ref{kpm.sec}), we take the system size $L=10^5$ with the expansion order $N_c=10^3$ in the Chebyshev polynomials
and we average over $10$ disorder realizations. Curiously the threshold values, $W_{t1}$ and $W_{t2}$ 
of Fig.~\ref{psi_mf} now show up as enhancements in $\rho_0$. This indicates that although every
pair of MFs are localized in two ends of the system, the proliferation
of states around the $E=0$ is causing indirect hybridization between them which then eventually displace
them from $E=0$. 
This is the DOS manifestation of the resilience threshold of MFs against the on-site disorder. 
Every pair of MFs are spatially pinned to the edge, and energetically pinned to $E=0$. This
pinning is protected by topology, as long as the disorder is not strong enough. 
But strong enough disorder cause depinning and breaks them pair by pair into the bulk of Anderson localized states. 

\subsection{Effect of disorder on WN}
To understand the meaning of the two thresholds beyond everyone of which a pair of MFs is lost,
we need to study the evolution of the topological number as a function of the disorder. 
As we will see, {\em at the resilience threshold, the WN of the system changes by one}. 
We use the method of Prodan \etal~\cite{Prodan-Non-commutative,prodanPRL-Mondragon-Shem2014} 
to calculate the WN for disordered systems (see appendix~\ref{wn.sec}). 
\begin{figure}[t]
    \centering
    \includegraphics[width=0.495\linewidth]{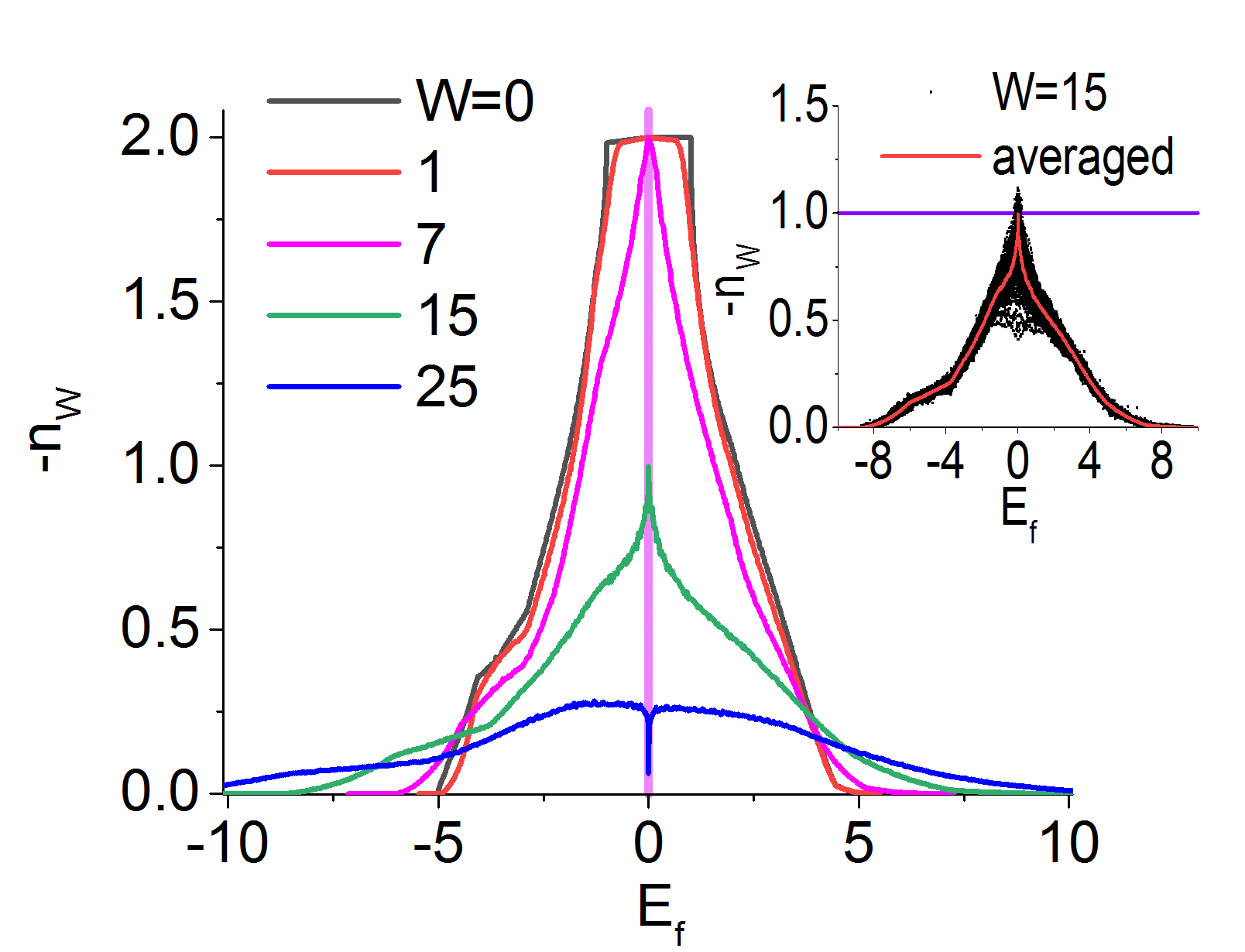}
    \includegraphics[width=0.495\linewidth]{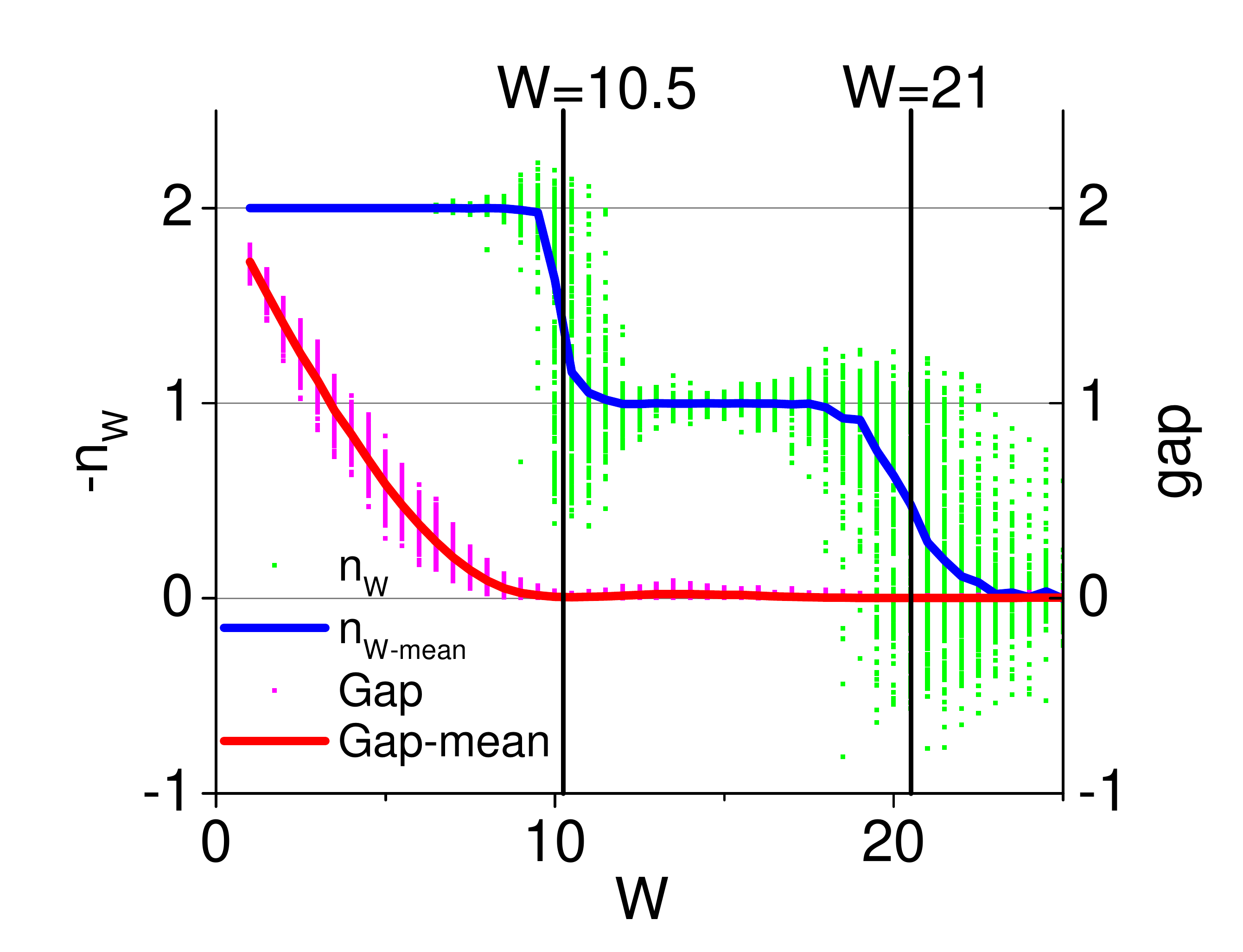}
    \caption{(left) The WN as a function of chemical potential, $E_F$ for various values of disorder $W$.  
    (inset) Averaging sharply pins $n_w$ of the half-filled system
to integer values. (right) WN at half-filling for $L=1000$ sites (blue) and the spectral gap (red)  
with $100$ configurations. The region between the first and second threshold is a TAI. 
}
    \label{fig:E-N}
\end{figure}
\begin{figure}[b]
    \centering
    \includegraphics[width=\linewidth]{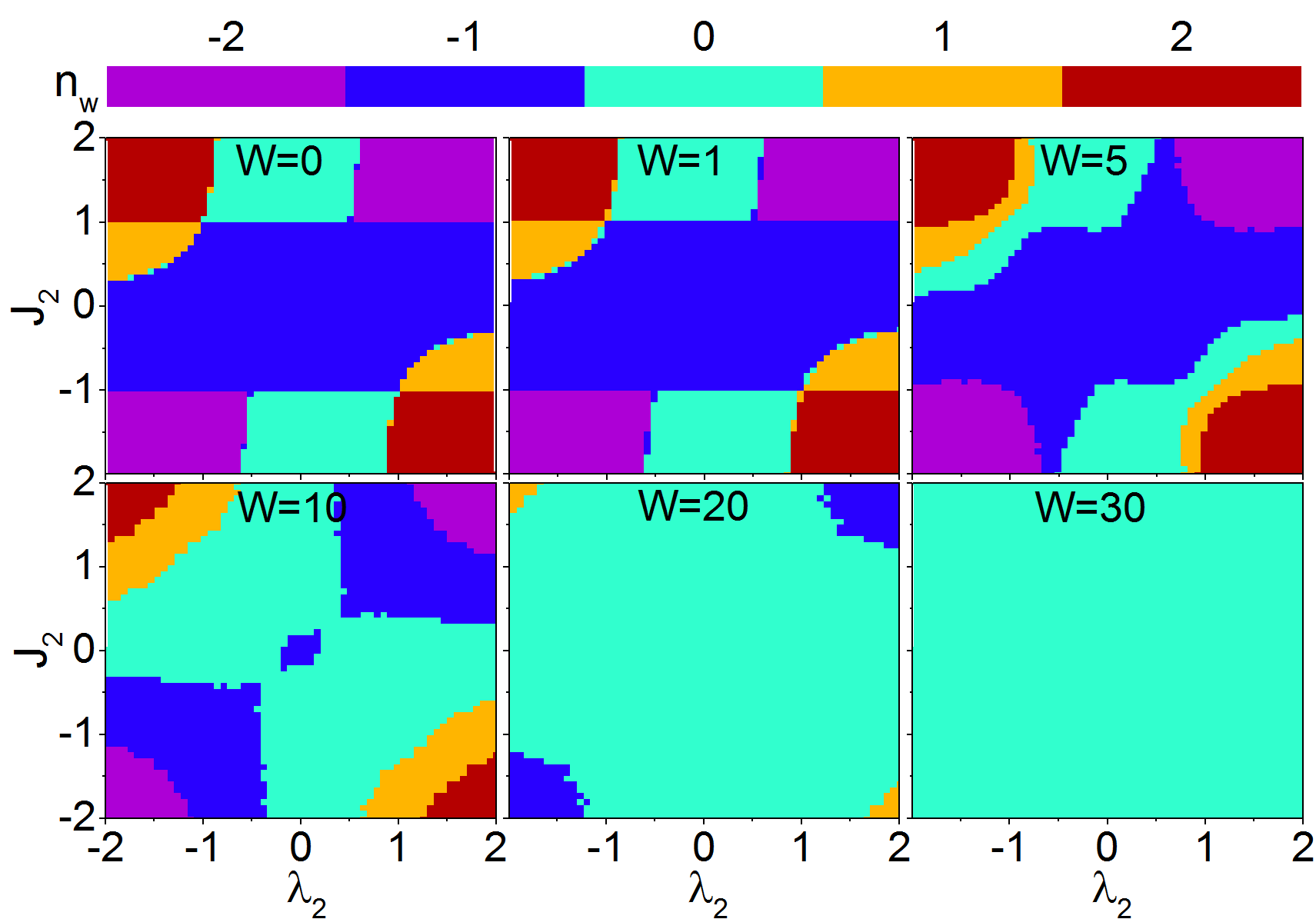}
    \caption{Topological phases of the disordered 2XY mode. The data are obtained for the system size of $L=1000$ with average over 
    $100$ configurations. 
    }
    \label{J-L2-winding}
\end{figure}
The WN of a system is obtained by summing over the occupied states. Therefore it crucially
depends on the position of the Fermi level $E_F$ that separates filled and empty single-particle states. 
As can be seen in Fig.~\ref{fig:E-N}-left, 
when the chemical potential crosses either of conduction or valence band, the
WN will not be an integer number. But when $E_F$ is in the middle of the spectrum, 
it is an integer. The inset indicates that averaging over disorder sharpens the WN. 
Our focus will be on the WN for $E_F=0$ which is shown in Fig.~\ref{fig:E-N}-right. 
As we increase $W$, there are two topological phase transitions across which WN changes as $-2\to -1\to 0$. Averaging over $10^2$
configurations produce smooth step-like functions. For larger systems, the steps get sharper. 
At the first topological phase transition at $W_{t1}\sim 10.5$ where two MFs across the ends of the system
annihilate each other (see Fig.~\ref{psi_mf}), the WN changes from $-2$ to $-1$, and simultaneously the spectral gap
entirely collapses. At this point the system is Anderson insulator, but still with the non-zero WN, $n_w=-1$. 
It is, therefore, qualified to be a TAI. This TAI phase persists until at a second threshold
value of $W_{t2}\sim21$ at which the remaining Majorana end modes annihilate each other by critically delocalizing over the entire
system. Beyond this point, the entire system is a trivial Anderson insulator. 

By repeating the above analysis for a range of values $(\lambda_2,J_2)$ we can map the
phase diagram of topological phases of the disordered 2XY model which is shown in 
Fig.~\ref{J-L2-winding}. In clean system~\cite{Jafari2016}, there are borders
across which the WN changes by $2$, as well as borders
across with the WN changes by $1$. Because of the one-by-one changing mechanism of the WN due to difference in the resilience of zero energy end modes, 
between each two phases having WN of $n_w$ and $n_w\pm2$, a nearby region with the WN $n_w\pm1$ penetrates. 
Eventually, at very strong values of disorder, the region with WN of zero conquers the entire phase diagram and the system will be a trivial Anderson localized insulator. 

Although the ultimate fate of the model is to end up in the topologically trivial state, there are 
certain topologically trivial regions in the clean system, e.g. region near $(\lambda_2,J_2)\approx(-1,2)$ 
which acquire a {\em disorder induced topology} as can be seen in $W=5$ panel of this figure. 
In the disordered system, the WN changes by $1$ across all borders. 
The regions of $n_w=\pm 1$ inserted by disorder between $n_w=\pm2$ and $n_w=0$ are TAI. 

\section{Resilience and Localization}
To investigate how the resilience thresholds show up in various indicators of Anderson localization,
we start by discussing inverse participation ratios (IPRs) which requires exact diagonalization and is numerically very costly. 
After corroborating against this method, we shall suggest an efficient and very fast algorithm to
precisely determine the disorder induced a change in topology. 
This will in the hindsight provide an alternative way of determination of the topological number
which applies equally well to clean and disordered systems. 
For a wave function $ \psi_{i,\lambda} $  at energy eigenvalue $ E_\lambda $, the IPRs is defined by,
$P(\lambda)={\sum_i{\left(\psi_{i,\lambda}^2 \right)^2}}/{\left(\sum_i{\psi_{i,\lambda}^2}\right)^2}$,
where $i$ denotes the lattice site. 
For extended states, it vanishes for large sizes as $1/N$. A plane wave is an extreme
example of this sort. The other extreme is a state sharply localized in one site for which $P=1$. 
Therefore for more localized states, $\lambda$ is closer to $1$ and for fully extended states, it is zero within ${\cal O}(1/N)$ accuracy.

\begin{figure}[t]
    \centering
    \includegraphics[width=0.50\linewidth]{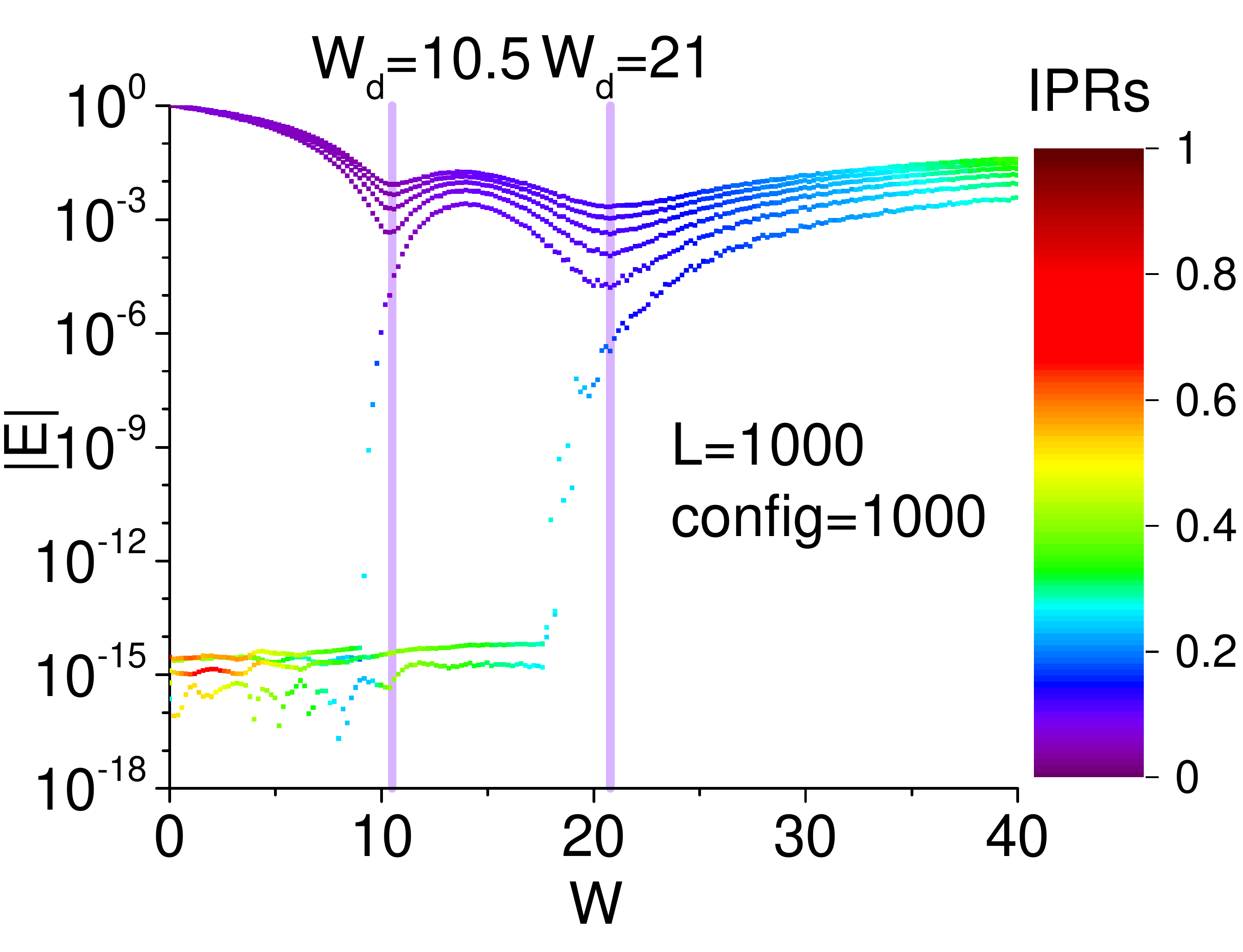}
    \includegraphics[width=0.49\linewidth]{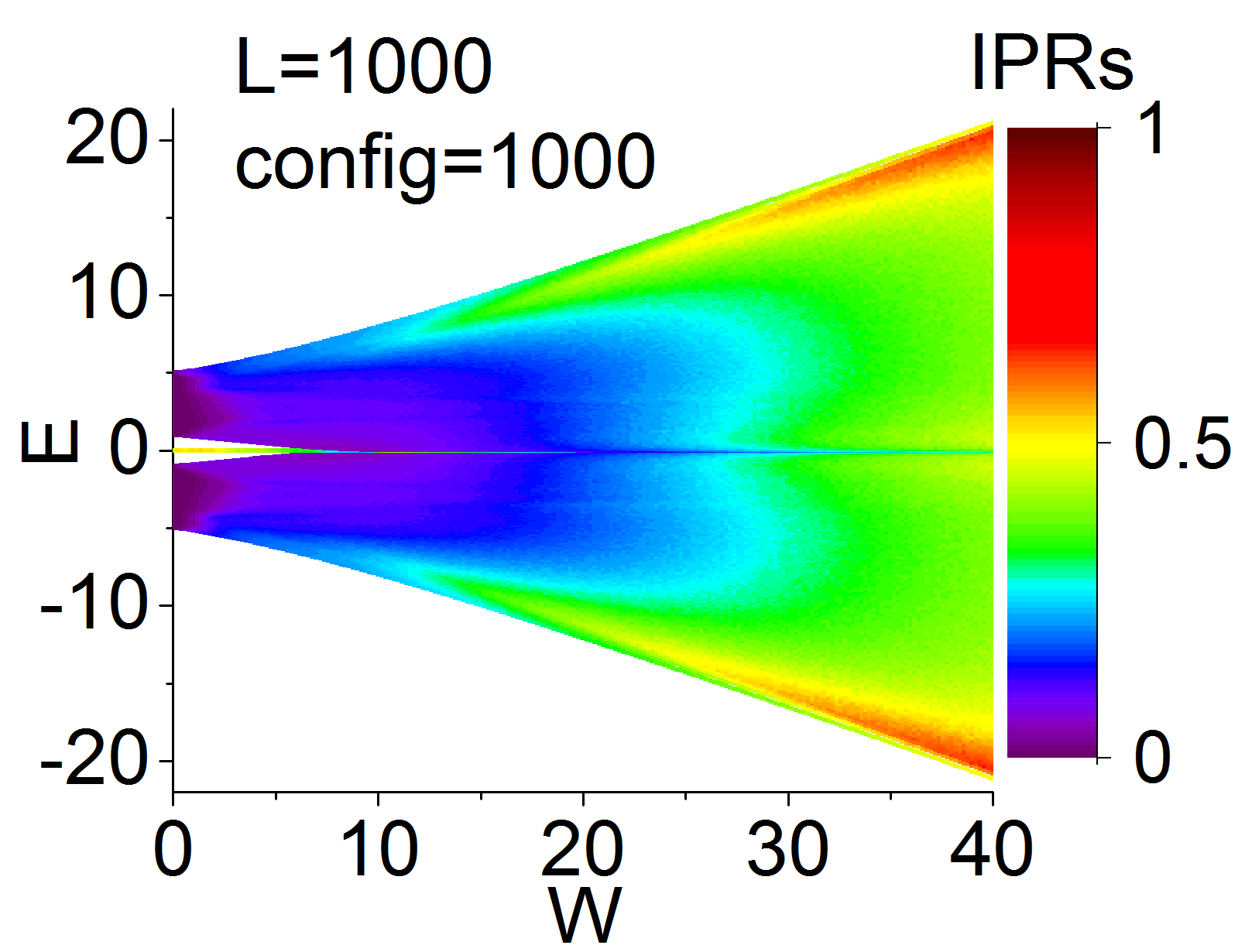} 
    \caption{(left) Tower of low-lying states as a function of disorder.
    The vertical axis is the absolute the value of energy $|E|$.
    The color code indicates the IPRs. (right)
    Intensity map of the average IPRs for the whole spectrum as a function of disorder strength. 
       }
    \label{fig:majorana-dis}
\end{figure}
Fig.~\ref{fig:majorana-dis}-left shows IPRs for low-lying parts of the tower of states as a function of disorder $W$. 
The vertical axis denotes the absolute value of energy, $|E|$. As can be seen by increasing disorder strength, 
states are pushed from both positive and negative side (due to particle-hole symmetry)
towards the $E=0$, and develop a dip at two threshold values. The energy of MFs 
is zero within the machine precision of $\sim 10^{-16}$. Near the threshold values 
$W_{t1}$ and $W_{t2}$ the MFs can not remain pinned to $E=0$ anymore. This
depinning tendency is signaled in the gradual reduction of IPRs which is encoded in the color code. 
Fig.~\ref{fig:majorana-dis}-right shows intensity map of IPRs for the entire spectrum of eigenvalues as a function of the disorder. 
By increasing the disorder, the spectral gap closes, and generically the IPRs increase, meaning that the states
tend to localize on fewer and fewer sites. The smallest amount of disorder is enough to localize the entire spectrum. 

\begin{figure}[b]
    \centering
    \includegraphics[width=0.495\linewidth]{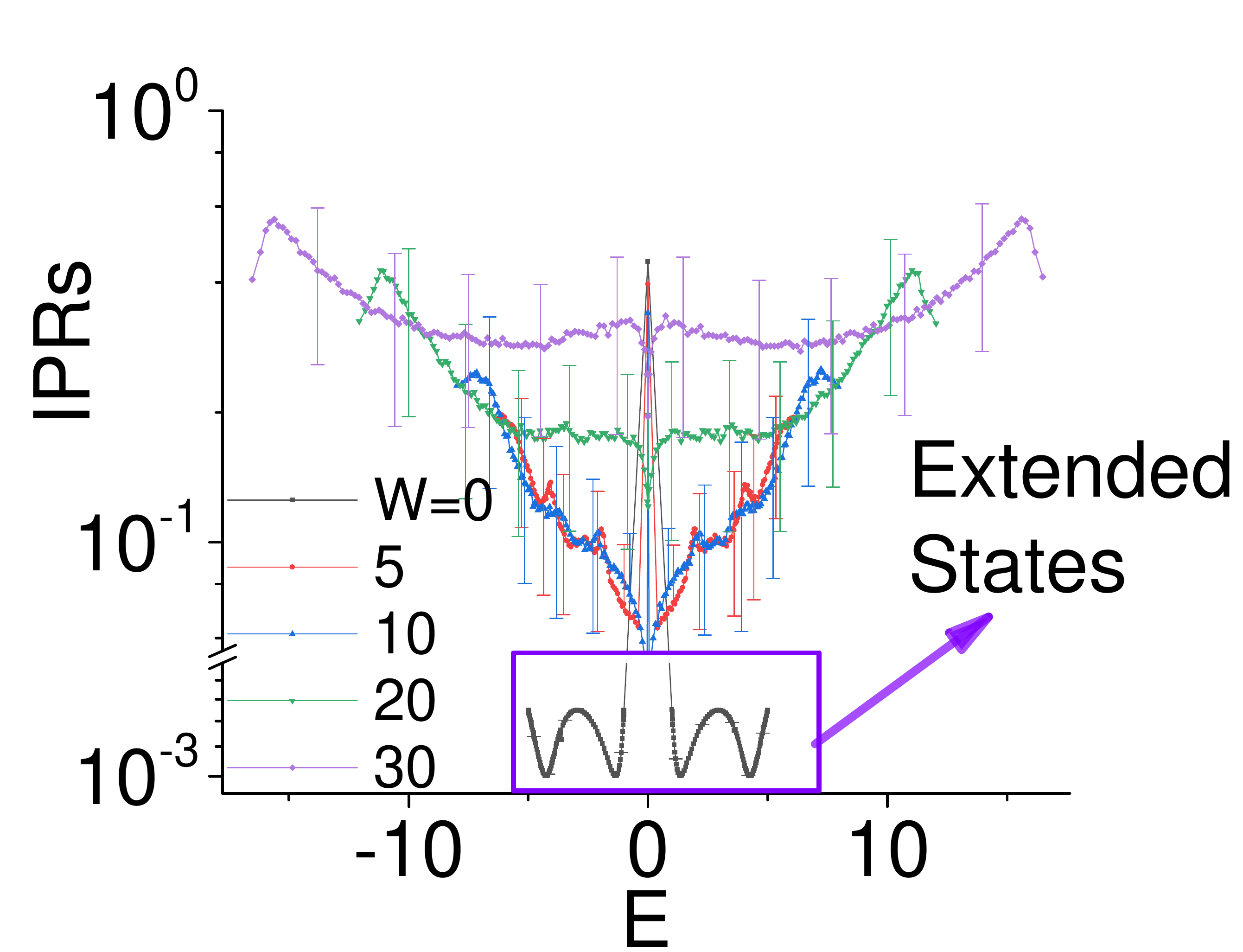} 
    \includegraphics[width=0.495\linewidth]{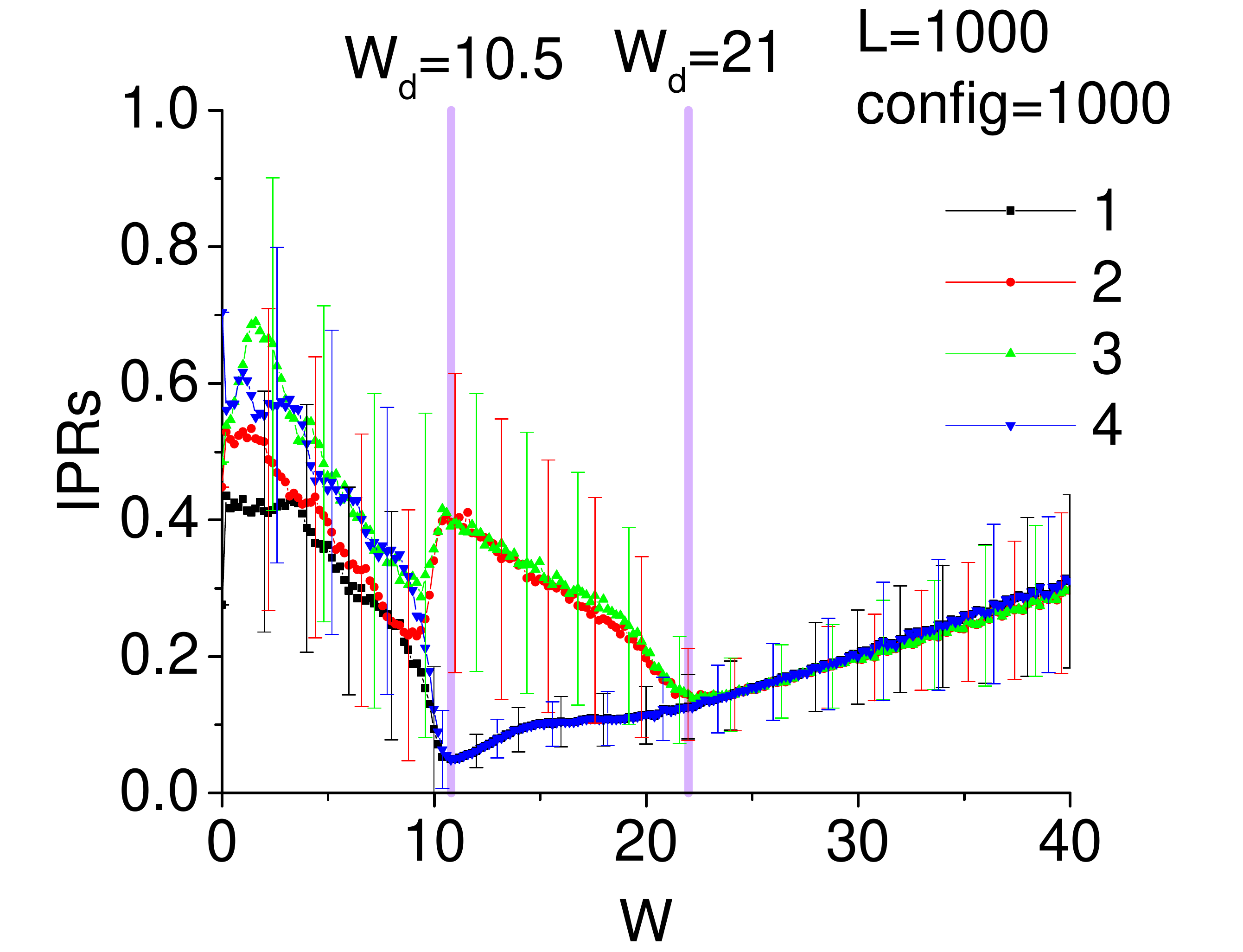} 
    \caption{
    (left) Constant disorder IPRs vs. $E$. MFs at zero energy are localized even for a clean system. 
    (right) IPRs for $E=0$ Majorana end modes. Resilience threshold is the turning point in the IPRs. 
}\label{IPRS}
\end{figure}
Fig.~\ref{IPRS}-left shows constant disorder cuts of the IPRs. For the clean system, the IPRs of the whole spectrum is
${\cal O}(1/N)$, except for the MFs that live at $E=0$. By adding disorder, they all become Anderson localized in the bulk. 
To understand the localization behavior of the Majorana zero modes, we pick four (two pairs of) MFs
and study their IPRs as a function of the disorder. This is the content of Fig.~\ref{IPRS}-right. 
Blue and black plots correspond to one pair of MFs, while red and green correspond to the other pair. 
For small values of disorder, the IPRs for
both pairs is a fraction of $1$, which indicates extreme localization. By increasing disorder, the IPRs tend to decrease, 
meaning that MFs tend to become less localized. At the first threshold, the IPRs of the first pair reaches a minimum, meaning that
they are stretched as much as possible. At the first threshold value, the first pair of MFs whose IPRs is reduced enough are drown into
the bulk. This is signaled by the upturn in their IPRs. Such a monotonic increase in IPRs as a function of the disorder is typical behavior of 
Anderson localized states.
Therefore at a first threshold, blue and black pair of MFs merge into the bulk of AL. 
The second pair (red and green curves) are however more resilient and refuse to delocalize further by
approaching the first threshold. Beyond the first threshold, $W_{t1}$ due to their Majorana character, their IPRs
unlike the Anderson localized states decreases by increasing $W$. Finally at a second threshold $W_{t2}$ the resilience of the last pair of 
MFs is exhausted and they surrender to disorder and merge into the bulk of Anderson localized states which is signaled as the upturn in this curve. 
Beyond the second threshold, there are no more MF pairs left and
hence all the knots of the wave function are opened to give a zero WN state. 
For disorder stronger than $W_{t2}$ all states show a generic Anderson localization behavior of increasing IPRs as a function of the disorder. 

The unusual (non-AL) decrease of IPRs of MFs gives them enough delocalization to let them
hybridize through higher-order processes within a nearly zero-energy subspace of the Hilbert space.
Increasing disorder enhances the density of nearly zero energy states and hence there will be enough density of very low-energy (localized) states that can mediate hybridization between the two Majorana partners at the two ends of the chain. In an empty lattice an IPRs of ${\cal O}(1/N)$ would be needed to qualify MFs
for critical delocalization over the entire system. In the disordered case, although the minimum
IPRs is an order of magnitude larger than a fully extended state, the enhancement of
the density of low-energy states (Fig.~\ref{fig:E-Dos}) allows them to efficiently hybridize and therefore depins them from $E=0$, ending their topological protection by saturating their
resilience threshold. 
It is curious to note in the right panel of Fig.~\ref{IPRS} that at the threshold values of disorder, 
not only the {\em average} IPRs develop a minimum,
but also the {\em fluctuations} of the IPRs are minimized. Understanding the fluctuations of
IPRs for topological and non-topological states deserves a separate investigation~\cite{Evers-Mirlin2000}.

So far we have found that the resilience threshold of Majorana fermions 
goes hand in hand with:
\begin{itemize}
   \item Change in the absolute value of WN: This is accompanied by wild spatial fluctuations of the WN
   \item Maximum in the density of zero-energy states as a function of disorder
   \item Minimal IPRs fluctuations
   \item Maximal extension of zero modes
\end{itemize}
The last fact allows us to use one of the powerful, precise, and fast techniques of Anderson localization
physics, namely the transfer matrix (TM) method to determine not only the resilience threshold, but also
by the above equivalence as an alternative tool to diagnose the topological index of the system. This is the subject of next section.

\section{Resilience and transfer matrix method}
So far we have identified the resilience threshold $W_t$ of MFs in various quantities. The TM method when appropriately modified to guarantee the convergence of the TM procedure, becomes a very cheap method to sharply determine $W_{t}$. This is the
subject of the present section. 
\begin{figure}
    \centering
    \includegraphics[width=0.8\linewidth]{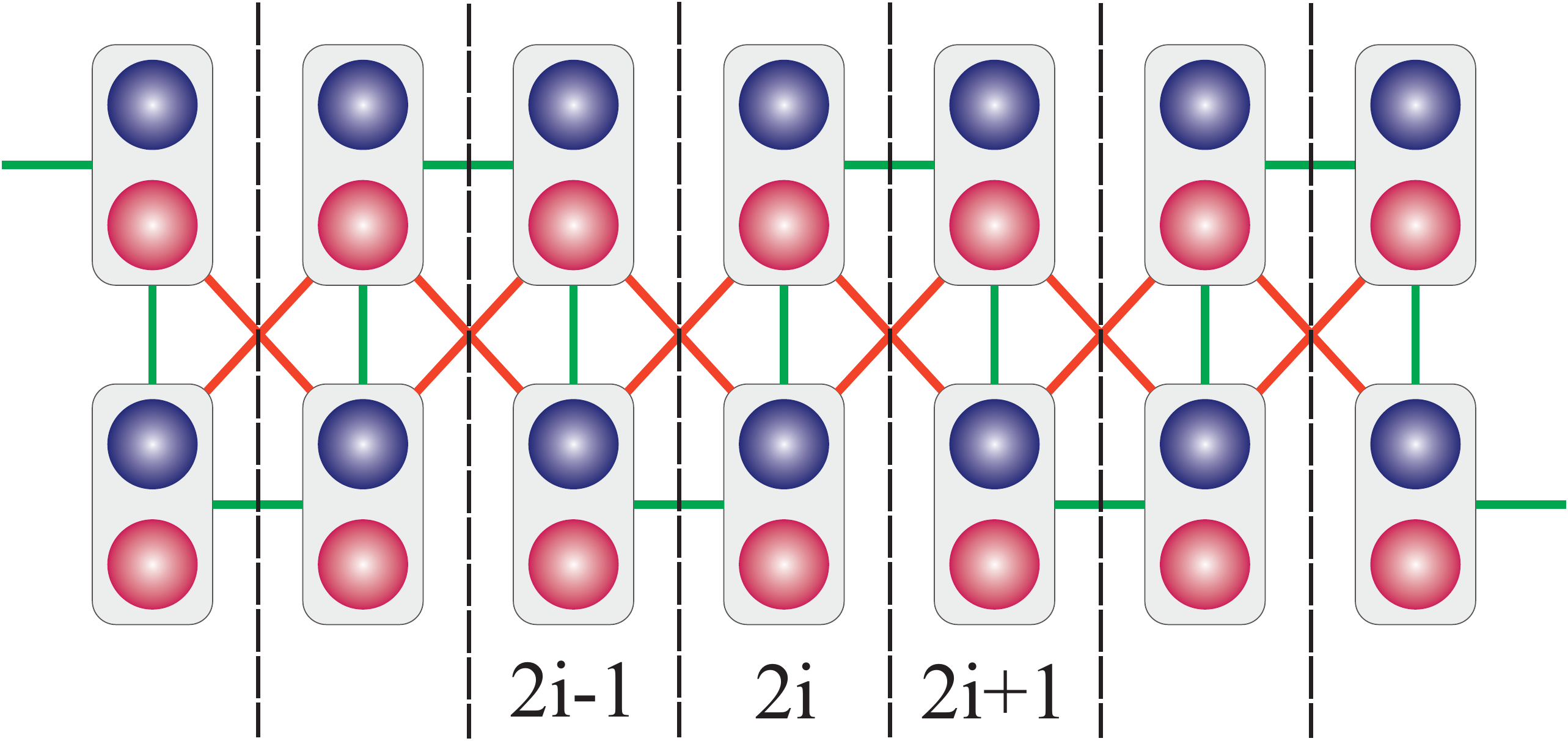}
    \caption{Transfer matrix for our Hamiltonian. Each slice contains two atoms. 
    Blue and red circles represent $\psi^e$ and $\psi^h$. Green lines and orange lines show nearest and next nearest neighbor
    couplings.}
    \label{fig:TM}
\end{figure}

    To calculate the localization length, we can use quasi-one-dimensional Schr\"odinger equation $H{\Psi }_i=E{\Psi }_i$\cite{MacKinnon1,MacKinnon2}. In our model, we need to calculate the localization length for the wave functions in the Nambu space. 
When we have next nearest neighbor, we are lead to organize the sites into the blocks depicted in Fig.~\ref{fig:TM} such that in the newly arranged form, the transfer of the amplitude of the wave function takes place only between neighboring blocks. In this basis, every block will have two sites
labeled by indices $1,2$ and the wave function $\Psi_i$ in the Nambu space will be 
$\Psi^T_i=\left(\psi^e_{i,1},\psi^h_{i,1},\psi^e_{i,2},\psi^h_{i,2}\right)$. 
This is effectively a four-channel quasi-one-dimensional problem. 
Within this representation, the wave equation becomes, 
\begin{align}
t_{i,i-1}{\Psi }_{i-1}+H_{i,i}{\Psi }_i+t_{i,i+1}{\Psi }_{i+1}=E {\Psi }_i,
\end{align}
which can be re-arranged to,
\begin{align}
\left(
\begin{array}{c}
{\Psi }_{i+1} \\
{\Psi }_{i} 
\end{array}
\right) =
T_{i+1,i}
\left(
\begin{array}{c}
{\Psi }_{i} \\
{\Psi }_{i-1} 
\end{array}\right),
\end{align}
where, 
\begin{align}
T_{i+1,i}=\left(
\begin{array}{cc}
t^{-1}_{i,i+1} \left(E-H_{i,i}\right) &- t^{-1}_{i,i+1} t_{i,i-1} \\
1 & 0 \\
\end{array}
\right).
\end{align}
    
    As can be seen in the Fig.~\ref{fig:TM} we have two types of slices labeled as $ 2i $ and $ 2i+1 $.   
The transfer matrix and onsite matrix for each slice are given by,
\begin{subequations}
    \begin{align}
    t_{2i,2i+1} =t_{2i+1,2i}^T &=\left(
    \begin{array}{cccc}
    0& 0 &j_2 & \lambda _2 \\
    0 & 0 &-\lambda _2 & -j_2  \\
    j_2 & \lambda _2 & j_1 & \lambda _1 \\
    -\lambda _2 & -j_2 & -\lambda _1 & -j_1 \\
    \end{array}
    \right),
    \\
    t_{2i+1,2i+2} =t_{2i+2,2i+1}^T&=\left(
    \begin{array}{cccc}
    j_1 & \lambda _1 &j_2 & \lambda _2 \\
    -\lambda _1 & -j_1  &-\lambda _2 & -j_2  \\
    j_2 & \lambda _2 & 0 &0\\
    -\lambda _2 & -j_2 &0 & 0\\
    \end{array}
    \right),
    \\
    H_{2i,2i}&=\left(
    \begin{array}{cccc}
    \epsilon _{i,1} & 0 & j_1 & \lambda _1 \\
    0 & -\epsilon _{i,1} & -\lambda _1 & -j_1 \\
    j_1 & -\lambda _1 & \epsilon _{i,2} & 0 \\
    \lambda _1 & -j_1 & 0 & -\epsilon _{i,2} \\
    \end{array}
    \right),
    \\
    H_{2i+1,2i+1}&=\left(
    \begin{array}{cccc}
    \epsilon _{i,1} & 0 & j_1 & -\lambda _1 \\
    0 & -\epsilon _{i,1} & \lambda _1 & -j_1 \\
    j_1 & \lambda _1 & \epsilon _{i,2} & 0 \\
    -\lambda _1 & -j_1 & 0 & -\epsilon _{i,2} \\
    \end{array}
    \right).
    \end{align}
\end{subequations}

\begin{figure}
    \centering
    \includegraphics[width=0.495\linewidth]{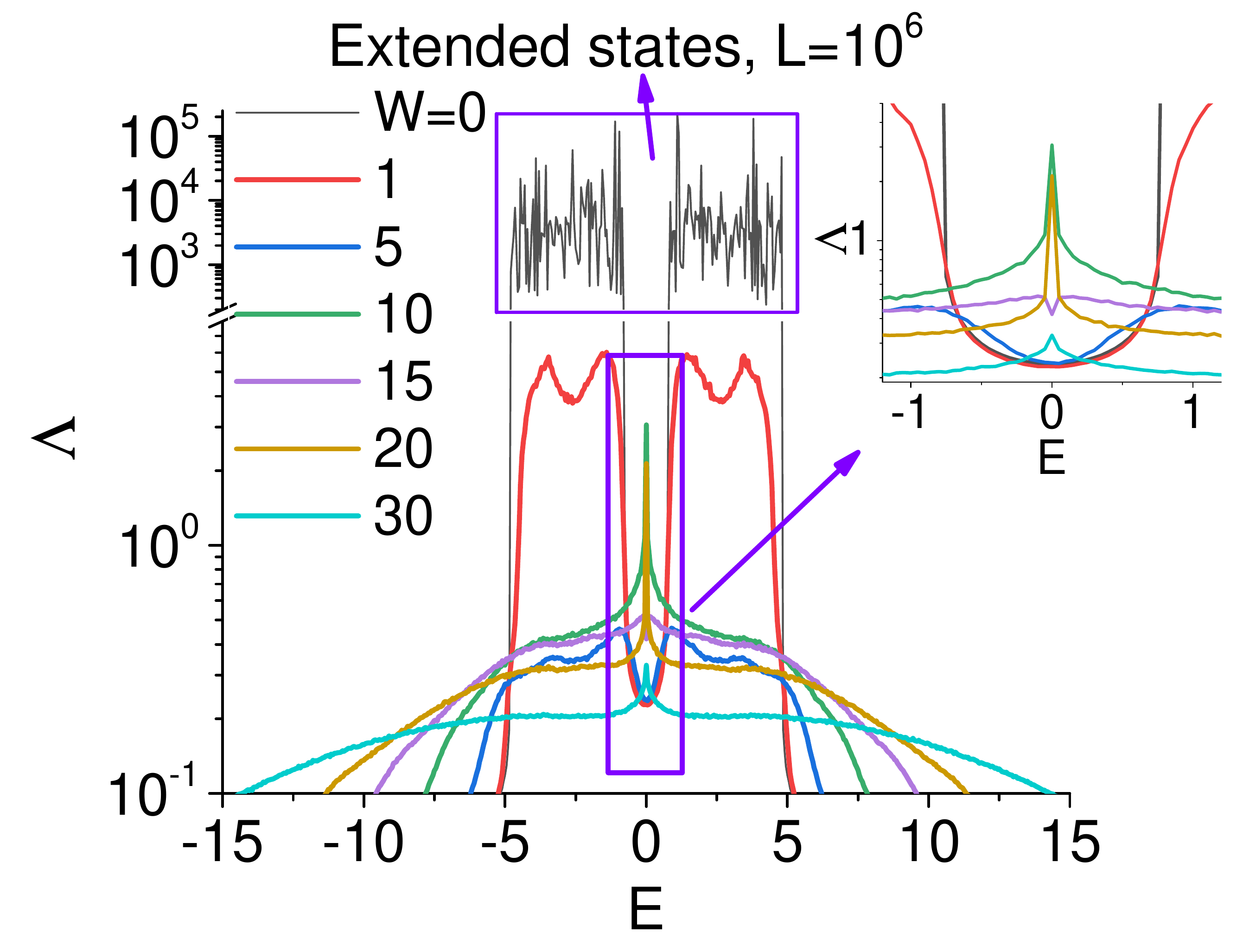}
    \includegraphics[width=0.495\linewidth]{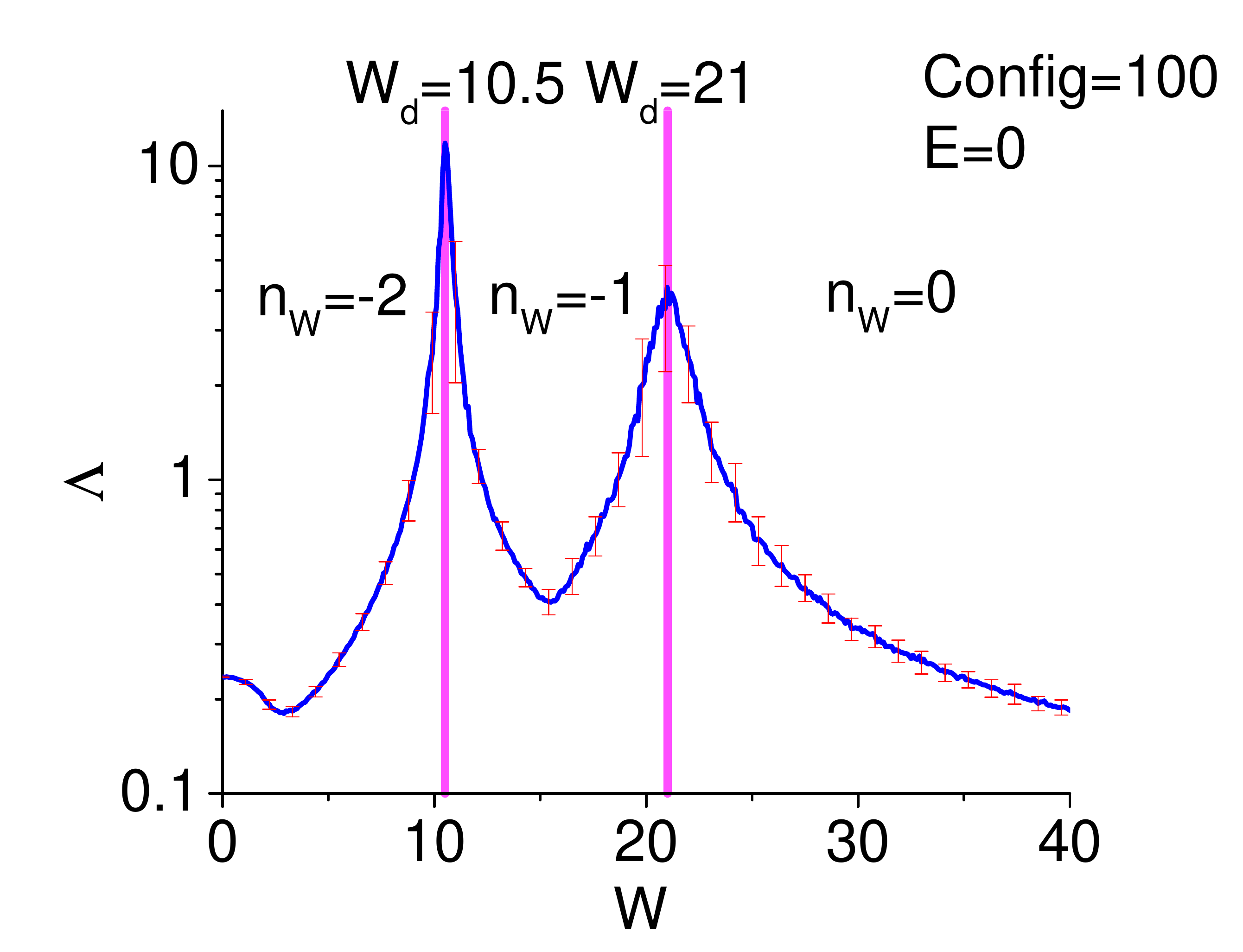}
    \caption{
    Localization length versus energy for various $W$ for (left) the entire energy spectrum
    -- The inset indicates that for $ W\approx 10 $ and $ W\approx 20 $ the localization length diverges at $ E=0 $ --
    and (right) $E=0$ only (i.e. MFs) as a function of disorder strength. 
    In Topological transition points $ W_{t1}= 10.5$ and $ W_{t2}=21 $ localization length of zero 
    energy states will diverge.}
    \label{fig:dis-L}
\end{figure}

    To calculate the localization length at the end of lattice, one needs to multiply the transfer matrices to form, 
    \begin{align}
    \Gamma=  \lim_{N \to \infty}{\left[\prod_{i=N,1}{T^\dagger_{i+1,i}\prod_{i=1,N}{T_{i+1,i}}}\right]^{1/{2N}}}
    \end{align}
    Diagonalizing the above matrix gives $i=1\ldots 8$ eigenvalues of the form $e^{\gamma_i}$.
    Due to the symplectic form of transfer matrix, the eigenvalues come in $e^{\pm \gamma}$ pairs. 
    The (largest) localization length $\Lambda$ is then calculated by,
    \begin{align}
    \Lambda=\frac{1}{|\gamma_{\rm min}|}
    \end{align}
    Details about obtaining the smallest positive Lyapunov exponent precisely can
    be found in Ref.~\onlinecite{MacKinnon1,MacKinnon2}.
    In our calculation, we choose $N$ large enough to guarantee the convergence of the localization length.  
   
In Fig.~\ref{fig:dis-L}-left we plot the localization length  $\Lambda$ as a function of energy $E$ of the states
for various values of disorder $W$. For $W=0$ all one-dimensional states, except the zero energy Majorana modes are extended. 
By adding the smallest amount of disorder, all bulk states become Anderson localized. By further increasing disorder (see the inset)
the localization length at zero energy tends to develop a peak near the threshold values of $W\approx 10,20$.

\begin{figure}[t]
    \centering
        \includegraphics[width=\linewidth]{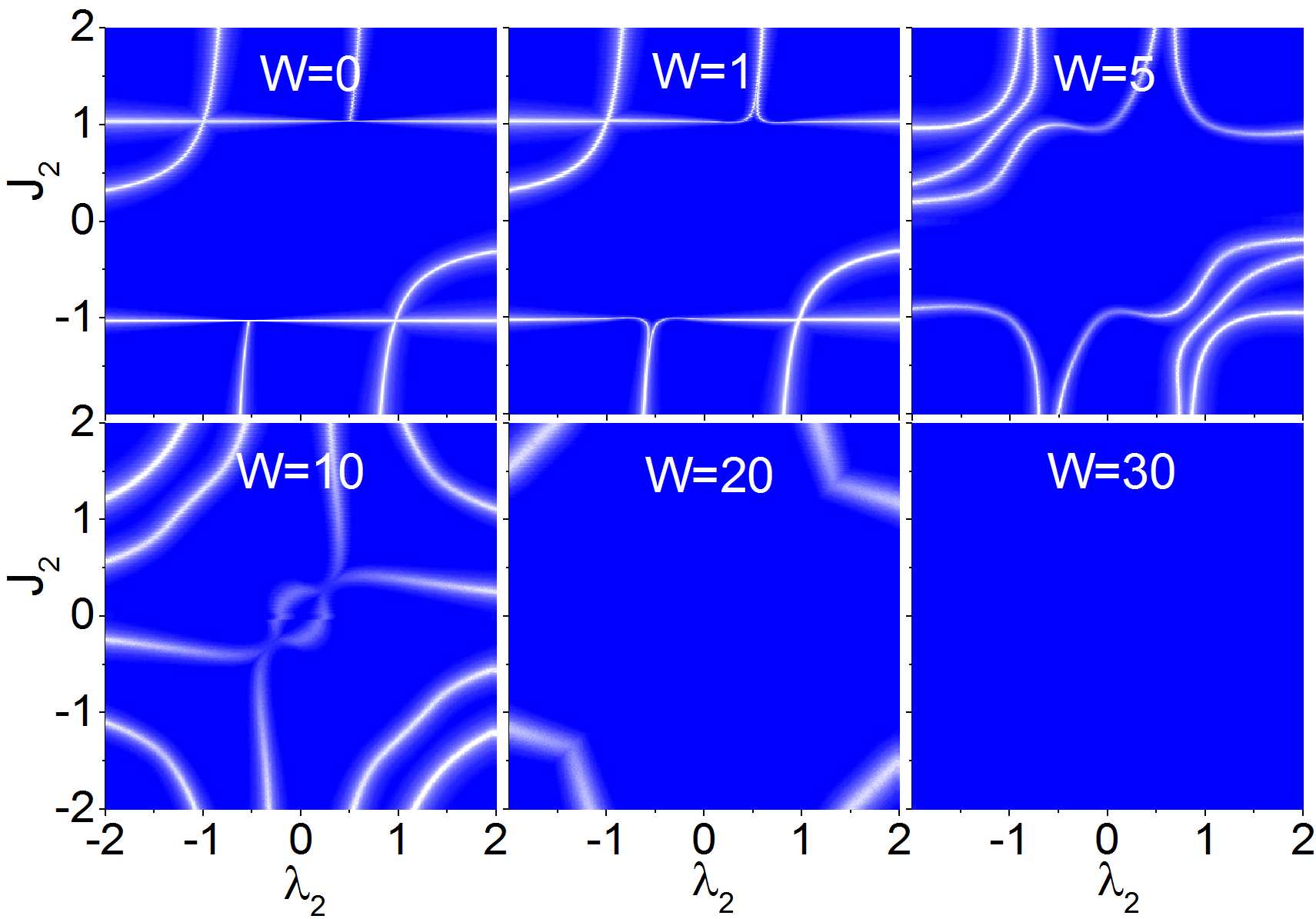}
    \caption{  Topological phase transitions from localization length calculated by transfer matrix. 
    Localization length diverges at transition points. }  
    \label{J-L2-Localizationlength}
\end{figure}

To see the critical delocalization of MFs at $W_{t}$, in Fig.~\ref{fig:dis-L}-right we focus on the calculation of localization length
for $E=0$ Majorana modes. This shows a very clear indication of a divergence of the localization length of Majorana zero modes at two threshold values indicated in the figure. The topological index (WN) is indicated in the figure. After each peak, the
absolute value of the WN decreases by one, and the system ultimately ends in a topologically trivial state with $n_w=0$ at
strongest disorder regime. Viewing this sequence of one-by-one change in the WN in reverse, elevates the TM method 
for a very quick method for the determination of $n_w$: 
The winding number can be determined by assigning $0$ to the most strongly disordered 
phase, and increasing the absolute value by $1$ upon crossing each divergence in 
$\Lambda$ of $E=0$ Majorana modes. The sign of the winding number is a matter of convention
and can be constructed from the symmetries of the Hamiltonian. 
Using the localization length $\lambda$ of Majorana zero modes, to map the phase diagram of the disordered 2XY model, we can sharply determine the phase boundaries. 
The phase boundaries are given in Fig.~\ref{J-L2-Localizationlength}. 
This figure agrees with Fig.~\ref{J-L2-winding} of the main text, but
it is more accurate and much more easier to determine with the transfer matrix method. 
Without knowledge of Fig.~\ref{J-L2-winding} the WN can be assigned to Fig.~\ref{J-L2-Localizationlength}
up to an overall sign ambiguity as explained above. 

The localization length as calculated from the Lyapunov exponent in the TM method provides not only a very precise 
but also numerically very economic (as it requires the TM procedure for $E=0$ energy only) determination of the 
resilience threshold of Majorana zero modes which also coincides with the onset of the change in the WN. 
The divergence of localization length at the resilience threshold furthermore
identifies the threshold of resilience with the maximal stretchability of MFs and hence substantiates the
claim of critical delocalization of MFs at the threshold values. 
Since the transfer matrix method is essentially free from finite size errors, as the amplitude of the wave function can be transferred
up to arbitrarily long distances, the system at the threshold values of the disorder must be scale invariant.
This entitles the disorder induced topological phase transitions to some sort of order parameter which 
can be encoded into appropriate nonlinear-sigma model in supersymmetry approaches~\cite{Altland2015}.

Note that the diverging length scale is a peculiar feature of MF zero modes. All other states are Anderson localized for
smallest disorder and therefore are left with no sense of the length scale of the entire system.

\section{Conclusion} 
We investigated the mechanism of change in the WN by the disorder. 
We established that every pair of MF have a threshold resilience at which they
critically delocalize and hence hybridize and depin from $E=0$. This is how they loose
their topological protection and become part of the Anderson localized bulk states. 
This explains why in the presence of disorder, across every boundary the WN changes by only one. 
After corroborating against costly and established methods of calculation of the topoloigcal index based on the polarization,
we showed that the above resilience threshold can be easily and precisely determined
by simply looking at the divergence of the localization length of {\em only} the $E=0$ states (corresponding to MF).
We curiously observed that at the threshold values, not only the IPRs of $E=0$ states reaches
a minimum, but the {\em fluctuations} of the IPRs are also suppressed.  
We furthermore showed that the resilience threshold is precisely where the WN changes by one. 
Given this, the localization length can be employed to sharply determine the winding number in
the BDI class of topological insulators. Note that this method makes no reference to the
polarization (related to Berry phase). 

It would be desirable to incorporate the above observations to construct an
effective order parameter theory to address the interplay between the
p-wave superconductivity and disorder~\cite{Narayanan2004}. 
The analogy with plateau-to-plateau transitions in quantum Hall effect and possible 
effective theories with theta term~\cite{Pruisken}
or possible RG interpretation~\cite{Jafari2017} is worth exploration.

\section{Acknowledgements}
We acknowledge discussions with A. Nersesyan and M. Dalmonte of the Abdus Salam ICTP.    

\appendix

    \section{Kernel Polynomial Method (KPM)}
    \label{kpm.sec}
    In KPM~\cite{Lee1,Amini-Jafari-2009,Habibi,Fehske} one expands the spectral functions in a set of orthonormal polynomials. The coefficients of expansion will be appropriate matrix elements or traces. 
    Then the traces can be stochastically calculated. Consider
    a Hamiltonian $H$ with energy eigenvalues $E$  in the range $[E_{\rm min},E_{\rm max}]$. To expand in Chebyshev polynomials,
    which are defined for arguments whose magnitude does not exceed $1$, one should first re-scale the Hamiltonian from 
    $H(E)$ to  $\hat{H}(\varepsilon)$ where $\hat{H}=(H-b)/a$ , 
    $\varepsilon=(E-b)/a$ , $b=(E_{\rm max}+E_{\rm min})/2 $ and  $a=(E_{\rm max}-E_{\rm min})/2$. 
    The normalized density of states can be expanded for range $\eps\in[-1,1]$ into Chebyshev polynomials as,
    \begin{eqnarray} 
    \hat{\rho}(\varepsilon)=\frac{1}{\pi \sqrt{1-\varepsilon^2}}\left (\mu_0 g_0+2\sum_{m=1}^{N_c}\mu_m g_m T_m(\varepsilon) \right)
    \label{dos} 
    \end{eqnarray} 
    where $T_m(\varepsilon)=\cos(m \arccos(\varepsilon))$  are the $m$th Chebyshev polynomials, 
    $\mu_m$ are Chebyshev moments and $g_m$ are the so called attenuation factors to minimize
    the Gibbs oscillations. $N_c$ is a cut off on the order of polynomials used in the expansion. 
    $\mu_m$  is defined as a trace formula, $\mu_m=1/r \sum_{r=1}^{M} \langle\phi_r\vert T_m(\hat{H}) \vert \phi_r \rangle$. 
    Since the trace does not depend on basis, we can choose 
    $\phi_r$ as random single-particle states and one should use $M$ as the number of random states used in the evaluation of the trace. 
    To obtain matrix elements of $ T_m(\hat{H})$ one can use recurrence relation of Chebyshev 
    polynomials, $T_m (\hat{H})= 2\hat{H} T_{m-1}(\hat{H}) - T_{m-2}(\hat{H})$ with 
    initial conditions $T_{1}(\hat{H})=\hat{H}$ and $T_{0}(\hat{H})=1$. This enables a computation of spectral functions
    without explicit diagonalization of the Hamiltonian. 

\section{Winding number}
\label{wn.sec}
This section is entirely based on Proden\cite{ Prodan-Non-commutative,prodanPRL-Mondragon-Shem2014}. 
The idea is to homotopically deform a given Hamiltonian $H$ to its flat band equivalent given by,
\begin{align}
    H\rightarrow Q=P_+-P_-
\end{align}
where $P_-$ and $P_+$ are projection operators to filled and empty bands, respectively, and are 
given by 
\begin{align}
    P_+=\sum_{E_n>E_f}|n\rangle \langle n|,~~~~~~P_-=1-P_+
\end{align}
Since in our model, chiral symmetry operator satisfies $S=S^\dagger$ and $S^2=1$, 
their eigenvalues are $\pm1$. Defining $S_\pm$ as projection operator to space of these eigenvalues, 
it can be represented as $S=S_+-S_-$. Every chiral symmetric operator, including $Q$ can be
represented as $Q=S_+ Q S_- + S_- Q S_+$ where 
$(S_\pm Q S_\mp)^\dagger = S_\mp Q S_\pm$ and $(S_\pm Q S_\mp)^{-1}= S_\mp Q S_\pm$. 
Then the WN can be calculated from the above canonical form using $Q_{+-}= S_+ Q S_-, \ (Q_{+-})^{-1}= S_- Q S_+=Q_{-+}$. 
Replacing $\partial_k$ with $-i[X,~]$ where $X$ is the position operator, and denoting the
summation over $k$ space degrees of freedom (per volume) by $\mathrm{tr}$, the WN in real space is given by
\begin{equation}\label{RWindingNr}
  n_w =  -\ \mathrm{tr}\big \{ Q_{-+} [X, Q_{+-}] \big \}.
\end{equation}
For a given realization of disorder, this formula allows the calculation of WN
in a single diagonalization procedure. Further averaging over disorder smoothens the variations
of the above WN.

\bibliographystyle{apsrev4-1}
\bibliography{Refs}

\end{document}